 \documentclass[prl,aps,showpacs,twocolumn]{revtex4}
\usepackage{amsmath,amsfonts,amssymb,graphics,graphicx,epsfig,color,times}
\usepackage[latin1]{inputenc}

\usepackage{cancel,ifthen,ulem}

\graphicspath{{images/}}
 
\DeclareMathSymbol{\epsilon}{\mathord}{letters}{"22}
\DeclareMathSymbol{\theta}{\mathord}{letters}{"23}
\DeclareMathSymbol{\rho}{\mathord}{letters}{"25}
\DeclareMathSymbol{\phi}{\mathord}{letters}{"27}
 
\DeclareMathSymbol{\varepsilon}{\mathord}{letters}{"0F}
\DeclareMathSymbol{\vartheta}{\mathord}{letters}{"12}
\DeclareMathSymbol{\varphi}{\mathord}{letters}{"1E}
\DeclareMathSymbol{\varrho}{\mathord}{letters}{"1A}

\begin{document}

 
\newcommand{\nn}{{\mathbbm{N}}}
\newcommand{\rr}{{\mathbbm{R}}}
\newcommand{\cc}{{\mathbbm{C}}}
\newcommand{\id}{{\sf 1 \hspace{-0.3ex} \rule{0.1ex}{1.52ex}\rule[-.01ex]{0.3ex}{0.1ex}}}
\newcommand{\me}{\mathrm{e}}
\newcommand{\mi}{\mathrm{i}}
\newcommand{\md}{\mathrm{d}}
\renewcommand{\vec}[1]{\text{\boldmath$#1$}}

\newcommand{\rB}{{\rm B}}
\newcommand{\rF}{{\rm F}}
\newcommand{\rDD}{{\rm 3D}}
\newcommand{\rED}{{\rm 1D}}

\newcommand{\cH}{\mathcal{H}}
\newcommand{\cO}{\mathcal{O}}
\newcommand{\cI}{\mathcal{I}}
\newcommand{\cZ}{\mathcal{Z}}

\newcommand{\hP}{\hat{\Psi}}
\newcommand{\hPd}{\hat{\Psi}^\dagger}
\newcommand{\hPe}{\hat{P}}
\newcommand{\hH}{\hat{H}}
\newcommand{\hA}{\hat{A}}
\newcommand{\hB}{\hat{B}}
\newcommand{\hX}{\hat{X}}
\newcommand{\hL}{\hat{L}}
\newcommand{\ha}{\hat{a}}
\newcommand{\had}{\hat{a}^\dagger}
\newcommand{\hc}{\hat{c}}
\newcommand{\hcd}{\hat{c}^\dagger}
\newcommand{\hr}{\hat{\rho}}
\newcommand{\hN}{\hat{N}}
\newcommand{\hn}{\hat{n}}
\newcommand{\hU}{\hat{U}}
\newcommand{\hV}{\hat{V}}
\newcommand{\hE}{\hat{E}}
\newcommand{\hS}{\hat{S}}
\newcommand{\hO}{\hat{O}}

\newcommand{\hcI}{\hat{\cI}}

\newcommand{\br}{{\bf r}}
\newcommand{\bv}{{\bf v}}
\newcommand{\bt}{{\bf t}}
\newcommand{\be}{{\bf e}}
\newcommand{\bF}{{\bf F}}

\newcommand{\scl}{\hspace{-0.6cm}}
\newcommand{\scls}{\hspace{-0.3cm}}

\newcommand{\bra}[1]{\langle #1 \vert}
\newcommand{\ket}[1]{\vert #1 \rangle}
\newcommand{\braket}[2]{\langle #1 \vert #2 \rangle}
\newcommand{\ko}[2]{\left[ #1, #2 \right]}
\newcommand{\ako}[2]{\left\{ #1, #2 \right\}}
\newcommand{\expv}[1]{\langle #1 \rangle}
\newcommand{\set}[1]{\left\{ #1 \right\}}

\newcommand{\hOg}{\hat{O}_{g}}
\newcommand{\Tr}{\mathrm{Tr}}
\newcommand{\eps}{\varepsilon}

\renewcommand{\H}{\mathcal{H}}
\newcommand{\ad}{\hat{a}^\dagger}
\renewcommand{\a}{\hat{a}}
\newcommand{\nb}{\hat{n}}
\newcommand{\cd}{\hat{c}^\dagger}
\renewcommand{\c}{\hat{c}}
\newcommand{\nf}{\hat{m}}
\newcommand{\f}{\hat{f}}
\newcommand{\NN}[1]{\sum_{\langle#1\rangle}}
\newcommand{\cred}{\color{red}}
\newcommand{\cblue}{\color{blue}}
\newcommand{\cgreen}{\color{green}}
\newcommand{\Uh}{\frac{U}{2}}
\newcommand{\trace}[2]{\textsf{Tr}_{#1}\left[#2\right]}


\newcommand{\kb}{k_{\text{B}}}
\newcommand{\Tc}{T_{\text{c}}}
\newcommand{\Eia}{E_{\text{ia}}}
\newcommand{\Ekin}{E_{\text{kin}}}
\newcommand{\Etrap}{E_{\text{trap}}}
\newcommand{\Hia}{\hH_{\text{ia}}}
\newcommand{\Hkin}{\hH_{\text{kin}}}
\newcommand{\Htrap}{\hH_{\text{trap}}}
\newcommand{\tint}{t_{\text{int}}}
\newcommand{\losc}{l_{\text{osc}}}
\newcommand{\telta}{\tilde{\delta}}

\title{Dynamics of pair correlations in the attractive Lieb-Liniger gas}
\author{Dominik Muth}
\email{muth@physik.uni-kl.de}
\affiliation{Fachbereich Physik und Forschungszentrum OPTIMAS, Technische Universit\"at Kaiserslautern, D-67663 Kaiserslautern, Germany}
\author{Michael Fleischhauer}
\affiliation{Fachbereich Physik und Forschungszentrum OPTIMAS, Technische Universit\"at Kaiserslautern, D-67663 Kaiserslautern, Germany}

\begin{abstract}
We investigate the dynamics of a 1D Bose gas after a quench from the Tonks-Girardeau regime to the regime of strong {\it attractive} interactions applying analytical techniques and exact numerical simulations. After the quench the system is found to be predominantly in an excited gas-like state, the so-called super-Tonks gas, however with a small coherent admixture of two-particle bound states. Despite its small amplitude, the latter component leads to a rather pronounced oscillation of the local density-density correlation with a frequency corresponding to the binding energy of the pair, making two-particle bound states observable in an experiment. Contributions from bound states with larger particle numbers are found to be negligible.
\end{abstract}
 
 
\date{\today}
 
\maketitle

Ultra-cold quantum gases in reduced spatial dimensions have attracted a lot of attention in
recent years \cite{Bloch2008}. This is because on one hand quantum effects play an increasing 
role in lower dimensions and on the other hand these systems became experimentally accessible using ultra-cold atomic gases.
A striking example for the pronounced effects is the effective fermionization of a one-dimensional Bose gas with repulsive interactions, described by the Lieb Liniger (LL) model \cite{Lieb1963a}, 
leading to the so-called Tonks-Girardeau (TG) gas \cite{Girardeau1960,Kinoshita2004,Paredes2004}. In the attractive case
the ground state of an $N$-particle LL gas is the highly localized McGuire cluster state \cite{McGuire1964}.
In the thermodynamic limit the gas is unstable, preventing direct experimental studies of
ground-state properties. However, in dynamical setups, attractive and repulsive gases are 
equally well accessible. A recent milestone in this direction is the creation 
of the super Tonks-Girardeau (sTG) gas \cite{Astrakharchik2005, Chen2010} by Haller et al. \cite{Haller2009} 
realized by a rapid sweep through a confinement induced resonance from the strongly repulsive to the strongly attractive 
side. The sTG gas is a highly excited, gas-like eigenstate, that does not contain any bound pairs or higher particle clusters. We here analyze the dynamics of this quench process by numerical simulations employing the time evolving block decimation (TEBD) \cite{Vidal2003,Vidal2004} algorithm recently applied to the simulation of the relaxation dynamics of the 
repulsive LL gas \cite{Muth2009} and a number of lattice models \cite{Gobert2005,Sirker2005,Flesch2008}. 


The Hamiltonian of a one-dimensional, trapped Bose gas with local interactions is given by
the Lieb-Liniger model \cite{Lieb1963a} with an additional potential term 
\begin{equation}\label{eq:H}
 \hH =  \int\!{\rm d}x\ \hat\Psi^\dagger(x)\biggl[\left(-\frac12\partial_x^2\right)
 + \frac{g}{2}\hPd(x)\hP(x) + V(x)\biggr]\hP(x)
\end{equation}
in units were $\hbar=m= 1$. $g$ is the interaction strength, which is related to the
1D scattering length $a_{\rm 1D}$ via $g=-2/a_{\rm 1D}$, and can have both signs. 
It is often characterized by the dimensionless Tonks parameter
$ \gamma = {g}/{\rho}$. $V(x)=\frac12\omega^2x^2$ describes an harmonic trap. 

The spectrum of ({\ref{eq:H}) is quite different depending on the sign of the interaction, however the positive- and negative-$\gamma$ spectra agree in the limits of weak {\it as well as} strong interactions \cite{Tempfli2008}. This can be most easily understood from the homogeneous problem of two particles with periodic boundary conditions (PBC): 
Fig.\ \ref{fig:2particlespectrum} shows the lowest lying states with vanishing center-of-mass momentum. The non interacting ground state (left {\it and} right end of the figure at energy $E=0$) has a constant relative wave function between the two particles. For finite interactions, the wave function must obey contact conditions, equivalent to the interaction Hamiltonian \cite{Lieb1963a}. It develops a peak at zero inter particle distance when an attractive interaction ($\gamma < 0$) is turned on and eventually forms a closely bound pair with binding energy $\gamma^2$ (see below). On the repulsive side ($\gamma > 0$) a dip-like kink emerges with increasing interaction, which eventually makes the wave function vanish at coinciding particle positions -- this is the famous fermionized TG gas. Most importantly when approaching the strong interaction regime
from the attractive side, the first excited state adiabatically connects to exactly the same fermionized state  -- the sTG gas. This matching continues for higher excited states and the scheme can be generalized to many particles, where the bound states can be classified by the number of dimers, trimers, etc. \cite{Muga1998}.

 \begin{figure}[htb]
   \epsfig{file=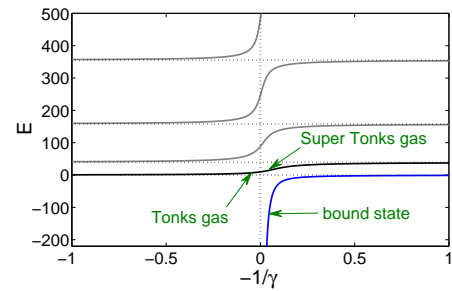,width=0.7\columnwidth}
    \caption{(Color online) Spectrum of LL Hamiltonian for two particles with vanishing total momentum 
on a ring of length $L=1$ as a function of inverse interaction strength. One recognizes equivalence 
of the spectra at vanishing ($|\gamma|\to 0$) as well as infinitely strong interactions ($|\gamma|\to\infty$).}
  \label{fig:2particlespectrum}
 \end{figure}

Fig.\ \ref{fig:2particlespectrum} indicates, that a quench from the TG regime to the strongly attractive regime will put the gas to good approximation in the sTG state. Experiments have successfully demonstrated this, while the difference between TG and sTG can be detected by their different compressibility \cite{Astrakharchik2005}.

We here consider a gas of $N$ particles confined by a harmonic trap initially being in the ground state for 
$\gamma =+\infty$. 
At $t=0$ the interactions are switched to the
strongly attractive side $\gamma\ll -1$. The trap plays a minor role, since interactions give the relevant time scale 
dominating over the kinetic energy resulting from the trap confinement.

We simulate the full many-body dynamics using an exact numerical technique. To this end the continuous model 
(\ref{eq:H}) is discretized, resulting in the sparsely filled Bose-Hubbard model \cite{Muth2010}. The lattice is actually finite, but comprises all of the gas (which does not expand globally on the time scale in question). The dynamics is then simulated using the TEBD scheme, employing a fourth order trotter decomposition \cite{Sornborger1999}. For the specific setup 
it is necessary that the conservation of the total particle number 
is taken into account explicitly. While time dependent simulations are generally limited to short times due to the linear growth of entanglement entropy \cite{Bravyi2006,Eisert2006}, this it not crucial here. Although we also observe such a linear growth, 
the increase is slow since we are close to an eigenstate. Thus we can go much beyond the time scale of interactions as in \cite{Muth2009} for as much as $N=18$ particles on a $1280$ sites lattice using a rather small matrix dimension of $\chi=100$ in the algorithm.

Fig.\ \ref{fig:g2nonlocal} shows the dynamical evolution of the density-density correlations, where we fix one position at the center of the cloud. Time is given in all figures in units of $4/\rho^2$. 
The initial state shows the typical feature of fermionization, i.e. $g^{(2)}(0,0)$ is zero and rises to one (no correlation) on a length scale proportional to the average inter-particle distance. In the limit $\gamma\to -\infty$
the correlations do not show much resolvable dynamics on the scale of the figure 
because the initial TG is close to the sTG state. However for the moderate interaction strength chosen, we see $g^{(2)}$ rising sharply around zero distance. This must be due to transitions to states other than the sTG state. 
One finds that the characteristic length scale of the peak at the origin is
given by the 1D scattering length $a_{\rm 1D}$. This indicates a finite admixture of the $N=2$ McGuire
cluster state. Note, that since $ \int {\rm d}x \expv{\hPd (x)\hPd (y)\hP(y)\hP(x)} = N-1$
the integral over $g^{(2)}(0, x)$ must be constant in time, as long as the density can be assumed to be homogeneous. The increase at $x_1-x_2=0$ must therefore be accompanied by a decrease at larger distances as seen in the correlation waves building up in Fig.\ \ref{fig:g2nonlocal}. 
The insert 
of Fig. \ref{fig:g2nonlocal} shows another interesting feature: Apart from small distances, where oscillations continue, the correlations become quickly stationary and show rather good agreement with $g^{(2)}(0,x)$ of a hard sphere Tonks gas (HS)
gas with hard-sphere radius $a_{\rm 1D}$. This verifies a recent finding by 
Girardeau and Astrakharchik \cite{Girardeau2010a}, who have shown that the
wave function of the sTG gas is identical to that of a TG gas where a small avoided volume
of size $a_{0}\approx a_{\rm 1D}$ is inserted into the two-particle wave
function around zero distance (HS Tonks gas).
In Fig.\ref{fig:g2nonlocal}, the density-density correlation  of
the HS Tonks gas is approximated by $g^{(2)}$ of a TG gas on a reduced volume where the
hard-sphere volume is excluded. This corresponds to the $t=0$ curve multiplied by $1-a_{\rm 1D}\rho$ and shifted by $a_{\rm 1D}$. 
For separations larger than  the
interparticle distance $1/\rho$ the non local nature of the TG - HS mapping \cite{Girardeau2010a}
\begin{equation}
\begin{split}
 &\Psi_{\rm HS}(x_1<x_2<\dots<x_N) \\&= \Psi_{\rm TG}\left(x_1, x_2+a_{0},\dots,x_N+(N-1)a_{0}\right)
\end{split}
\end{equation}
prevents however a simple calculation of $g^{(2)}(0,x)$.

 \begin{figure}[htb]
   \epsfig{file=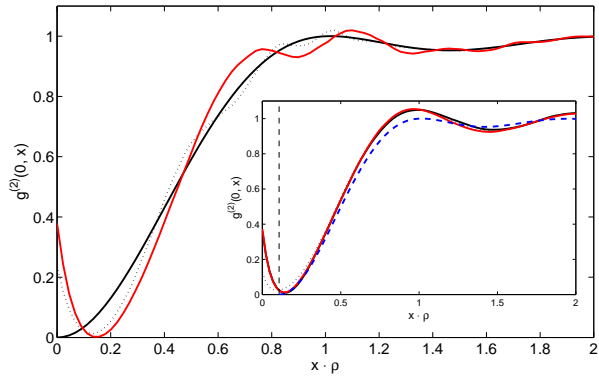,width=.95\columnwidth}
    \caption{(Color online) Time evolution of the non-local density-density correlation 
function for interaction strength $\gamma=-18.7931$ calculated for $N=18$ particles using TEBD. Curves are shown for 
$0$ (black),  $1/4$ (dotted), and $1/2$ (red, maximum value at $x=0$) oscillation periods of the local correlation. {\it insert}: $g^{(2)}$ at times of $3.5$ (black), 4 (dotted), and $4.5$ (red) periods. The vertical dashed line indicates $x=a_{\rm 1D}$. The blue dashed line shows an approximation to the HS ground state mentioned in the text.}
  \label{fig:g2nonlocal}
 \end{figure}

Fig.\ \ref{fig:g2local}a shows $g^{(2)}$ for equal positions as a function of time for various values of the interaction strength. 
 The correlation function grows as a power law with exponent depending on the interaction strength. By 
a linear fit to the numerical data, we 
find it growing from $1$ in the free case to a value of about $4/3$ in the strongly attractive case. 
We see that $g^{(2)}$ rises up to a finite value much smaller than $1$ for reasonably strong interaction. This reflects the fact, that most of the gas ends up in the fermionized sTG state. For longer times and stronger interactions, we observe however a
rather peculiar oscillatory behavior with large modulation depth. This can only be understood as a result of an
interference between two components of a coherent superposition, the sTG state 
and a bound state. In fact
the oscillation frequency coincides with  the binding energy
of a  {\it pair} of particles in the McGuire state.  Thus the 
dynamics is strongly affected by the contribution of bound pairs. Moreover, there is no sign of a relaxation, 
as observed in the repulsive case \cite{Muth2009}.
In Fig.\ \ref{fig:g2local}b the local three-particle correlation $g^{(3)}$ is plotted. One recognizes
that $g^{(3)}$ remains extremely small, showing that higher-order cluster states are not formed in the 
interaction quench. This agrees well with the finding in \cite{Girardeau2010a} where the overlap of the TG
wave function with the McGuire cluster state was calculated. 

 \begin{figure}[htb]
   \epsfig{file=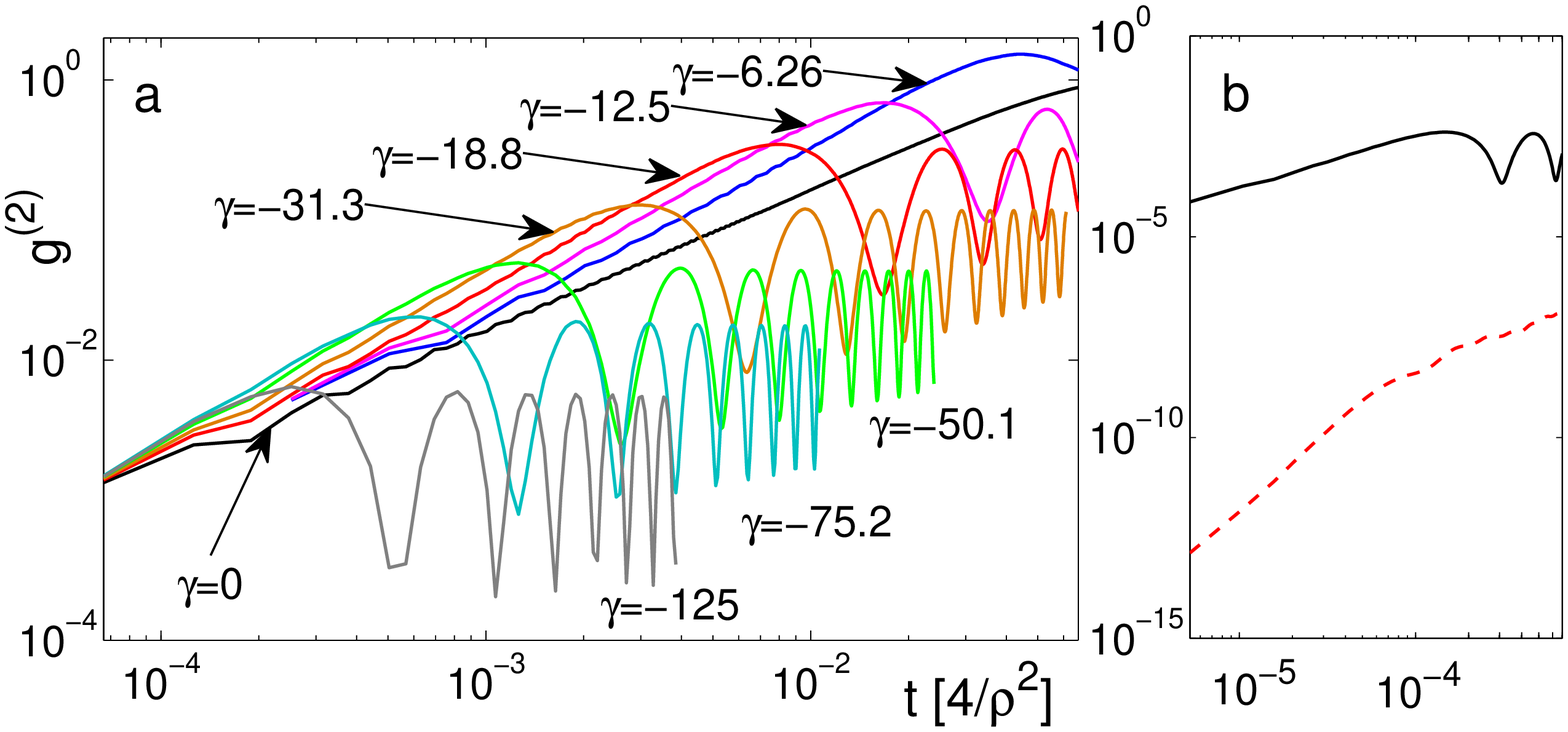,width=\columnwidth}
    \caption{(Color online) Time evolution of the local two particle density density correlation in a system of $N=18$ 
particles calculated via TEBD (colored lines). gray line: two particle case with PBC and $\gamma=0$.
{\it insert:} Time evolution of local three-particle correlation $g^{(3)}$ (red dashed) at the trap center and $g^{(2)}$ (black) for comparison at $\gamma=-145$. (The artifacts for very short times are due to the finite-size time steps
used by the numerical algorithm.)}
  \label{fig:g2local}
 \end{figure}

We will show now that the two-particle correlations in the gas can be very well reproduced by a system containing only 
$N=2$ particles. This is due to the fact that for strong interactions, eigenstates are well approximated by pair product states of the Jastrow-Bijl type \cite{Girardeau2010a}
\begin{equation}
\Phi(x_1,\dots,x_N) = \left[\prod_{ i< j} \phi(x_i-x_j)\right] \prod_{j=1}^N
\exp\left(-\frac{x_j^2}{2}\right).
\end{equation}
We impose periodic boundary conditions ($L=1$), which is reasonable for the comparison to the trapped gas, since it will be homogeneous in good approximation over the length scale of some inter-particle distances. The PBC problem gives analytical expressions and allows to extract the scaling of certain quantities with $\gamma$ in the strongly interacting regime. This problem has been solved for attractive interactions in the appendix of the original paper by Lieb and Liniger \cite{Lieb1963a}. We will use their solution in the following.

The Hamiltonian for the two particle problem reads in first quantization
$\hH = -\frac12\left(\partial_1^2+\partial_2^2\right) + g\delta(x_1-x_2)$.
All eigenstates of the LL model can be constructed from coordinate Bethe ansatz \cite{Yang1969}. 
In the primary sector ($0\le x_1\le x_2\le L$), the solution is $\phi(y=x_1-x_2) = 2A e^{i\frac{\delta}{4}}\cos\left(\frac{\delta}{2}(y-\frac12)\right)$.
%
%
$A$ is an interaction-dependent normalization constant, and $\delta$ is related to the scattering
phase shift $\Theta=-2\tan^{-1}((k_2-k_1)/2\gamma)$ via
$\Theta = \delta/2 - \pi$. Note that $e^{i\frac{\delta}{4}}$ is not a simple phase factor 
as $\delta$ will be imaginary for the two-particle bound state.

We will now calculate asymptotic expressions for the bound state $\phi_{\text b}$, where $\gamma \rightarrow - \infty$, as well as TG and sTG states $\phi_\pm$, where $\gamma \rightarrow \pm \infty$.
For the bound state  $\phi_{\text b}$ we need to find an imaginary solution of the Bethe equation
%
 ${\delta}/{2\gamma} = {1}/{\tan(\delta/4)}.$
%
Any $\delta$ that is not purely real, must be purely imaginary, as $\gamma$ is real. Substituting $\delta = i\telta$ we find in the strongly interacting limit $\telta = (-2\gamma)$.
With this we calculate the normalization of the wave function, yielding 
$A_{\text b}\, \to\, \sqrt{{\telta}/{2}}$.
Thus the wave function at coinciding particle positions reads
\begin{equation}
 \phi_{\text b}(0, 0) = 2A_{\text b}e^{-\frac{\telta}{4}}\cosh\frac{\telta}{4} \xrightarrow{\gamma\rightarrow -\infty}
-\sqrt{-\gamma}.
\end{equation}
As the density $\rho$ is 2 everywhere, this results in  $g_{{\text b}}^{(2)} 
\rightarrow\, {\telta}/{4}=-{\gamma}/{2}$.
%
%

We denote the lowest lying gas like states $\phi_{\pm}$ for $\gamma\rightarrow\pm\infty$. From the Bethe
equation we see that a real solution $\delta$ will be close to $2\pi$. Expanding the tangent around its singularity at $\pi/2$ we get
$\delta = 2\pi\left(1-\frac{2}{\gamma}\right).$
For the normalization this means $A_{\pm}\, \to\,  {1}/{\sqrt{2}}$,
such that the wave function at coinciding particle positions becomes
\begin{equation}\label{eq:chizerozero}
 \phi_\pm(0, 0) = 2A_\pm e^{\frac{i\delta}{4}}\cos\frac{\delta}{4} 
\xrightarrow{\gamma\rightarrow -\infty} i\sqrt{2}\frac{\pi}{\gamma}.
\end{equation}
The local two-particle correlation is in this case $g_{\pm}^{(2)} = {\pi^2}/{\gamma^2}$.
%
%
The $\gamma^{-2}$ scaling is well known \cite{Kheruntsyan2003} and is the same as in the many particle case.

The overlap between the initial TG gas state $\phi_0 = \lim_{\gamma\rightarrow\infty}\phi_+$ and the bound state for finite $\gamma$ can easily be calculated. In the strongly interacting regime one finds $\eps \equiv
\braket{\phi_0}{\phi_b} \, \to\, -2\sqrt{2}\pi\gamma^{-3/2}$.

 \begin{figure}[htb]
   \epsfig{file=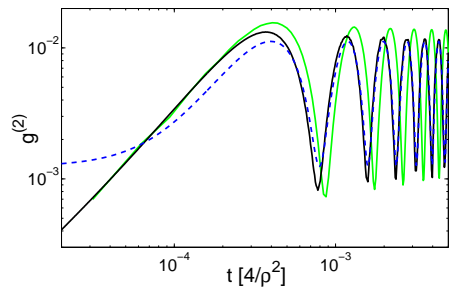,width=0.7\columnwidth}
    \caption{(Color online) Comparison of the many body results with the results from the two particle system 
with PBC in the case $\gamma=-89.0355$. The thick black line corresponds to the two particle case, the thin green one shows the $N=9$ particle case (calculated via TEBD), and the blue dashed one is the beating approximation (\ref{eq:twoa}).}
  \label{fig:2particlevstebd}
 \end{figure}

We now want to calculate the local correlation dynamics in the two particle case. Numerically this is easily doable by constructing all eigenstates solving the Bethe equation and re-expressing the initial state using this eigenbasis.  On the other hand from the above calculations we can derive simple approximations for the strongly interacting regime which are very good and give insight into the nature of the observed oscillations.
We can decompose the initial state $\ket{\phi_0}$ according to
\begin{equation}
 \ket{\phi_0} = \eps\ket{\phi_{\text b}} + \Bigl(\ket{\phi_0}-\eps
\ket{\phi_{\text b}}\Bigr) = \eps\ket{\phi_{\text b}}+\ket{\bar\phi_0} .
\end{equation}
Note that for large $\vert\gamma\vert$, $\ket{\bar\phi_0}$ is approximately normalized. 
One finds that for $t=0$ the wave function at coinciding particle positions is given by
\begin{equation}\label{eq:barzerozero}
 \bar\phi_0(0,0, t=0)\, \xrightarrow{\gamma\rightarrow -\infty}\,  i2\sqrt{2}\frac{\pi}{\gamma}.
\end{equation}
$\bar\phi_0$ is however not an eigenstate, but is composed out of low lying gas like states, that have an energy 
spread much smaller than the binding energy of the pair state. For small times one can ignore the energy differences
and thus ignore the time dependence of $\bar\phi_0$. This results in the correlation
\begin{equation}\label{eq:singlea}
 g^{(2)}(t)\, \xrightarrow{\gamma\rightarrow -\infty}\, 8\frac{\pi^2}{\gamma^2}\Bigl[1-\cos(\gamma^2t)\Bigr].
\end{equation}
This expression describes the initial increase of $g^{(2)}$ as observed in the many-particle calculation
(and in the exact solution of the $N=2$ case) very well. It does predict, however, oscillations with unity modulation depth, which is not true for the exact solution. The reason for this is that $\bar \phi_0$ contains in addition to the dominant, lowest gas-like state (i.e. the sTG state) $\phi_-$ also small admixtures of higher lying gas states which
oscillate in time all with slightly different frequencies. For larger times these oscillations 
lead to an effective dephasing in the interference part of $g^{(2)}$. 
On the other hand the direct contribution of these higher excited gas states to $g^{(2)}$ is negligible.
Thus an approximation which is much better suited to describe the large
time behavior is $\ket{\phi_0} \approx \eps\ket{\phi_{\text b}} + \ket{\phi_-}$.
Comparing (\ref{eq:chizerozero}) and (\ref{eq:barzerozero}) shows, that we have only changed a factor of $2$ such that
\begin{equation}\label{eq:twoa}
 g^{(2)}(t)\, \xrightarrow{\gamma\rightarrow -\infty}\, \left\{5-4\cos\Bigl[
\left(\gamma^2+\pi^2\right)t\Bigr]\right\}
\frac{\pi^2}{\gamma^2},
\end{equation}
where we used that the sTG gas energy is $\pi^2$ for strong interaction, giving a minor correction to the frequency. 
On short timescales this expression is of course invalid. However at times $t>1/\gamma^2$ it becomes a much better approximation than (\ref{eq:singlea}) as shown in Fig.\ \ref{fig:2particlevstebd}.

In summary we have shown by numerical TEBD simulations that an interaction quench of a 1D Bose gas from strong repulsive to strong attractive interactions puts the gas predominantly into the lowest gas-like excited state, the sTG gas. There is however a small coherent admixture of two-particle bound states that results in a large amplitude oscillation
of the local density-density correlation with a frequency corresponding to the energy difference between sTG gas and
bound pair state. At the same time higher-order correlations 
remain extremely small showing that more deeply bound, multi-particle cluster states are not formed in the quench.
Analytical calculations of the $N=2$ case where shown to reproduce the results
of the many-particle simulations with high accuracy. This indicates that the many-body
state can be well approximated by a Jastrow-Bijl type pair product wave function, where each term is a coherent
superposition of a gas-like state with a very small component of a two-particle bound state.
The peculiar oscillations of $g^{(2)}$ show furthermore that despite their small weight, the two-particle cluster states are accessible to experimental probes. For strong interactions, the bound pairs are highly co-localized. Since in all physical realizations of the LL model, the true inter particle potential is of finite range, details of the potential will 
show up in the binding energy. In this way, the two particle correlation dynamics can be used to measure details of the underlying true potential. In order to assess whether the effect predicted is accessible in current
experiments let us estimate the beat frequency in SI units. Asymptotically it is given by $\omega = {\gamma^2\hbar\rho^2}/{4m}$ with $m$ being the mass of the particle. This translates into $\omega = {\gamma^2  N 
\omega_\parallel }/4$, where $\omega_\parallel$ is the longitudinal trap frequency, i.e., for the experiment presented in \cite{Haller2009} the beating frequency is of the order of several ${\rm kHz}$. This
is comparable to the transversal trap frequency and thus an experimental observation requires either to go to smaller
values of $\gamma$ or to use tighter traps. Finally our paper also shows that the TEBD algorithm is suitable for the simulation of dynamical processes in strongly interacting, continuous quantum gases.

We are indebted to Hanns-Christoph N\"agerl and Elmar Haller for valuable discussions. This work was 
supported in part by the SFB TRR49 of the Deutsche Forschungsgemeinschaft and the
graduate school of excellence MAINZ/MATCOR.



\end{document}